\documentclass[sigconf]{acmart}
\AtBeginDocument{%
  }

\setcopyright{acmlicensed}
\copyrightyear{2025}
\acmYear{2025}
\acmDOI{preprint}
\acmConference[Preprint]{preprint}{XXX}

\begin{document}

\title[Health Dialogue Authoring Tool]{HealthDial: A No-Code LLM-Assisted Dialogue Authoring Tool for Healthcare Virtual Agents}

\author{Farnaz Nouraei}
\email{nouraei.f@northeastern.edu}
\orcid{0000-0002-0406-4410}
\affiliation{%
  \institution{Northeastern University}
  \city{Boston}
  \state{MA}
  \country{USA}
}

\author{Zhuorui Yong}
\email{yong.zh@northeastern.edu}
\affiliation{%
  \institution{Northeastern University}
  \city{Boston}
  \state{MA}
  \country{USA}
}

\author{Timothy Bickmore}
\email{t.bickmore@northeastern.edu}
\affiliation{%
  \institution{Northeastern University}
  \city{Boston}
  \state{MA}
  \country{USA}
}

\renewcommand{\shortauthors}{Nouraei et al.}

\begin{abstract}
  We introduce HealthDial, a dialogue authoring tool that helps healthcare providers and educators create virtual agents that deliver health education and counseling to patients over multiple conversations. HealthDial leverages large language models (LLMs) to automatically create an initial session-based plan and conversations for each session using text-based patient health education materials as input. Authored dialogue is output in the form of finite state machines for virtual agent delivery so that all content can be validated and no unsafe advice is provided resulting from LLM hallucinations. LLM-drafted dialogue structure and language can be edited by the author in a no-code user interface to ensure validity and optimize clarity and impact. We conducted a feasibility and usability study with counselors and students to test our approach with an authoring task for cancer screening education. Participants used HealthDial and then tested their resulting dialogue by interacting with a 3D-animated virtual agent delivering the dialogue. Through participants’ evaluations of the task experience and final dialogues, we show that HealthDial provides a promising first step for counselors to ensure full coverage of their health education materials, while creating understandable and actionable virtual agent dialogue with patients.

\end{abstract}


\begin{teaserfigure}
  \includegraphics[width=\textwidth]{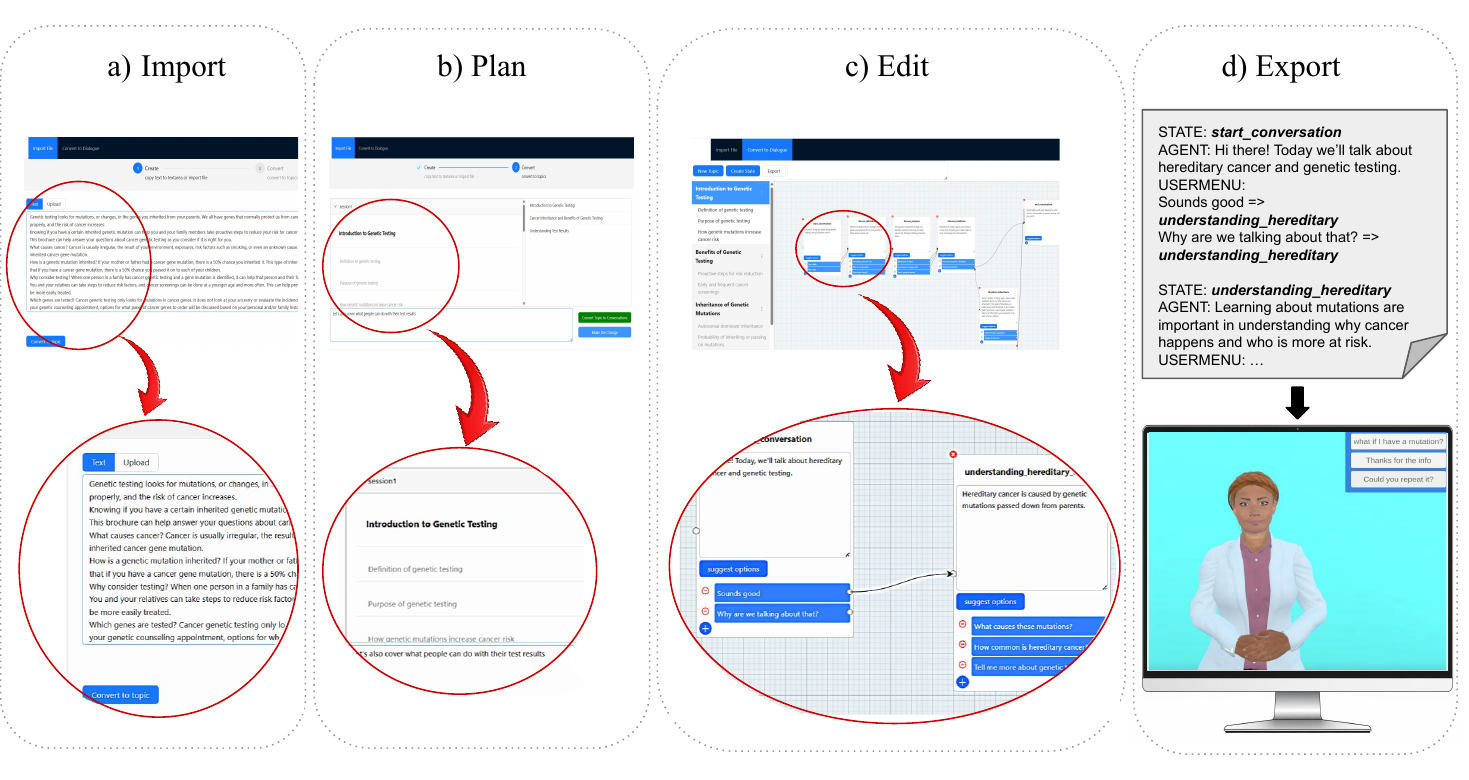}
  \caption{HealthDial Interaction Flow. a) Health educator begins by importing materials they would like to teach patients. b) HealthDial  suggests the number of virtual agent dialog sessions to cover the material, and a topic title for each session. The user can prompt HealthDial to make changes in this session plan. The session plan is then converted into a series of agent conversations.  c) The conversation is visualized using a finite state machine, showing the flow of the dialogue for each session. The user can flexibly add or remove states, edit agent utterances, add or change patient response options, and connect states. d) Once complete, each session conversation is exported in a format that can drive a virtual agent.  }
  \Description{figure 1}
  \label{fig:teaser}
\end{teaserfigure}

\received{20 February 2007}
\received[revised]{12 March 2009}
\received[accepted]{5 June 2009}

\maketitle

\section{Introduction}
The global healthcare landscape faces persistent and growing challenges in resource allocation, provider burnout, and the reach of high-quality patient education. Shortages of skilled healthcare professionals are especially severe in domains that require nuanced, repeated communication with patients \cite{world2016global, harp2023shortage}. As demand for health education outpaces the availability of seasoned providers, there is increased pressure to develop scalable strategies for training and supporting healthcare staff—many of whom may not receive adequate training in conversational skills necessary for effective patient engagement \cite{mata2021training}.

One promising avenue for augmenting provider capacity and reducing burnout is the use of conversational virtual agents. When appropriately designed, virtual agents have been shown to effectively deliver health education, support health behavior change, and even scale the reach of interventions by enabling remote and asynchronous interactions \cite{wolfe2015efficacy, bickmore2013randomized}. Face-to-face interventions utilizing virtual agents have demonstrated notable success in counseling for substance use \cite{yasavur2014let, steenstra2024virtual, lisetti2013can}, genetic risk communication \cite{zhou2022virtual, zhou2021automating, visvanathan2023engage}, among other issues. Virtual agents are particularly well suited for educational topics with a clear structure and informational objectives. An example of these topics is hereditary cancer screening or routine vaccination—for which pamphlets are traditionally used, but interactive dialogue holds greater potential for patient comprehension and action \cite{visvanathan2023engage}.

Despite their promise, virtual agents are difficult for non-technical healthcare providers to create due to the technical requirements of designing dialogue and system deployment. Thus, close collaborations between health experts and computer scientists are often needed to allow for creation of these agents, which involves health experts either role-playing and recording health sessions, or drafting textual “scripts” of health conversations, which are then manually converted into system-readable formats of dialogue by computer scientists to deploy in virtual agents. However, this process can be cumbersome and error-prone for both teams, as it is limited by each party’s understanding of each other’s subject matter expertise.

Moreover, healthcare virtual agents are usually created using predetermined dialogue trees or graphs, where a set of response options are available to the patient following each agent utterance \cite{laranjo2018conversational, bickmore2010maintaining}. The task of planning an entire educational conversation—anticipating patient concerns and preparing nuanced responses—can itself be a cognitive hurdle for clinicians \cite{debowski2023design, islam2023design}. Providing complete coverage of patient educational needs requires not only content expertise, but also fluency in the iterative, branching structure of natural dialogue. Studies in counselor training have long recognized that even scripted or role-played dialogues struggle to capture the full diversity of real patient inquiries \cite{pilnick2018using}, often resulting in incomplete or overly rigid conversations when used for virtual agent design. In addition, health counseling interventions are often long and spread across multiple sessions \cite{bickmore2005establishing}, which renders the design activity even more difficult for providers.

In this work, we aim to shift this paradigm by introducing HealthDial, a novel no-code dialogue authoring tool for healthcare providers and educators. HealthDial enables users to create session-based conversational flows by simply uploading or manually entering standard patient education materials (e.g., pamphlets or guidelines). Leveraging recent advances in large language models (LLMs), HealthDial automatically generates an initial, structured conversation plan covering key topics and typical patient utterances over multiple sessions, and presents it as an editable finite state machine (FSM) in an intuitive visual interface. The final dialogue can be exported in text format and in the form of a hierarchical transition network similar to \cite{bickmore2010maintaining}, which can be used in virtual agent systems.

By introducing a no-code, LLM-assisted approach to health dialogue authoring, we aim to:
1) reduce the technical barrier of writing system-readable scripts, allowing providers to visually and directly author, review, and revise each dialogue branch,
2) assist providers in planning and scaffolding complex health conversations by offering high-quality, context-appropriate LLM-generated drafts, and 
3) ensure safety and reliability by outputting deterministic dialogue for virtual agent delivery, and allowing the provider to edit LLM generations to prevent inappropriate or inaccurate model outputs from reaching patient interactions, thus addressing a critical limitation of directly deploying artificial intelligence (AI) in live settings \cite{bickmore2021mitigating}.

To evaluate this approach, we conducted a feasibility and usability study in the context of hereditary cancer screening education. Participants—including counselors and students with previous patient-facing experience—used an educational pamphlet, along with the HealthDial system, to author virtual agent dialogues on this topic. Participants subsequently tested these dialogues by interacting with a virtual agent. We found that through a usable and intuitive interface, HealthDial allowed participants to translate static guidance into actionable, patient-centered virtual counseling experiences. Importantly, the LLM suggestions were generally trusted and accepted, and were deemed to be aligned with the counselors’ own best practices. By supporting a human-in-the-loop authoring paradigm, our system lays the groundwork for broader adoption of conversational agents in health education, while maintaining the oversight and adaptability that only skilled healthcare experts can provide.

\section{Related Work}
\subsection{Virtual agents in healthcare} \label{subsec:rwhealth}
Virtual agents have demonstrated substantial promise in healthcare, offering scalable means for patient education, counseling, and behavior change support. Numerous studies highlight positive outcomes in diverse domains, such as chronic disease management \cite{griffin2021conversational}, substance use counseling \cite{li2024virtual, lisetti2013can}, and genetic risk communication \cite{fallah2024investigating}. When designing these agents, unique requirements in healthcare—such as patient safety, validity, and complete coverage of educational content—set a high bar for their dialogue design. Often, this is done through domain experts outlining of educational content and anticipation of patient responses \cite{rossen2012crowdsourcing}. Unfortunately, this approach does not guarantee the generalization of resulting dialogues to various situations or full coverage of conversation possibilities \cite{reiter2003acquiring}. The safety concerns in healthcare virtual agents often results in manual creation and editing of dialogue \cite{denecke2023framework}, rather than using automated approaches, making it a cumbersome and error-prone process. HealthDial aims to bridge these gaps by providing a platform that leverages large language models to scaffold the domain experts' authoring processes by providing an initial dialogue draft to save time, while allowing a manual editing process to ensure safety.

\subsection{Computer-Based Dialogue Authoring Tools}
General commercial tools (e.g., DialogFlow, Microsoft Bot Framework) focus on short, transactional dialogues through intent and entity mapping. These platforms, though often user-friendly, may not be adapted for the requirements of applications such as health counseling, where multi-turn, session-based conversations need scripting by domain experts, and safety demands the avoidance of unpredictable, unconstrained model outputs \cite{bickmore2010maintaining, beinema2022wool, bickmore2021mitigating}. Academic and open-source tools such as WOOL have focused on bridging these gaps, offering user-friendly interfaces and scripting capabilities for multi-session coaching and health advice. WOOL’s visual editor and scripting language are designed to support collaboration between technical and health experts, providing support for multi-session scenarios and extensible dialogue functionalities \cite{beinema2022wool}. 

In another line of work, crowdsourcing techniques have been used that allow domain experts to leverage crowdworker suggestions, to iteratively refine the conversation utterances and response options shown within the agent interactions, thus addressing the knowledge generalization issue discussed in Section \ref{subsec:rwhealth} \cite{rossen2012crowdsourcing, choi2020leveraging}. Similarly, tools such as those developed for the Atlas-Andes system facilitate rapid authoring by providing graphical user interfaces and knowledge acquisition pipelines for non-technical users, focusing on reducing the bottlenecks in scripting complex, branching educational dialogues for virtual agents \cite{jordan2001tools}. However, these approaches generally require authors to write structured dialogue scripts, with limited scaffolding to assist topic planning or conversational coverage, thus imposing significant design and cognitive loads on healthcare providers.

While prior work has addressed several challenges in dialogue authoring, there remains a lack of tools that are simultaneously user-friendly for non-technical healthcare professionals, robust against AI hallucinations, and capable of automatically scaffolding multi-session health counseling dialogues with a wide coverage of patient response options. HealthDial advances dialogue authoring by combining the power of large language models with a deterministic, finite state machine scripting environment and a no-code visual editor, tailored specifically for healthcare virtual agent dialogue creation. This human-in-the-loop LLM-based approach directly responds to the gaps in previous authoring systems and the healthcare-specific need for safe, transparent, and comprehensive agent-driven patient education.

\section{HealthDial System}
\subsection{System Design}
\begin{figure}
  \centering
  \includegraphics[width=\linewidth]{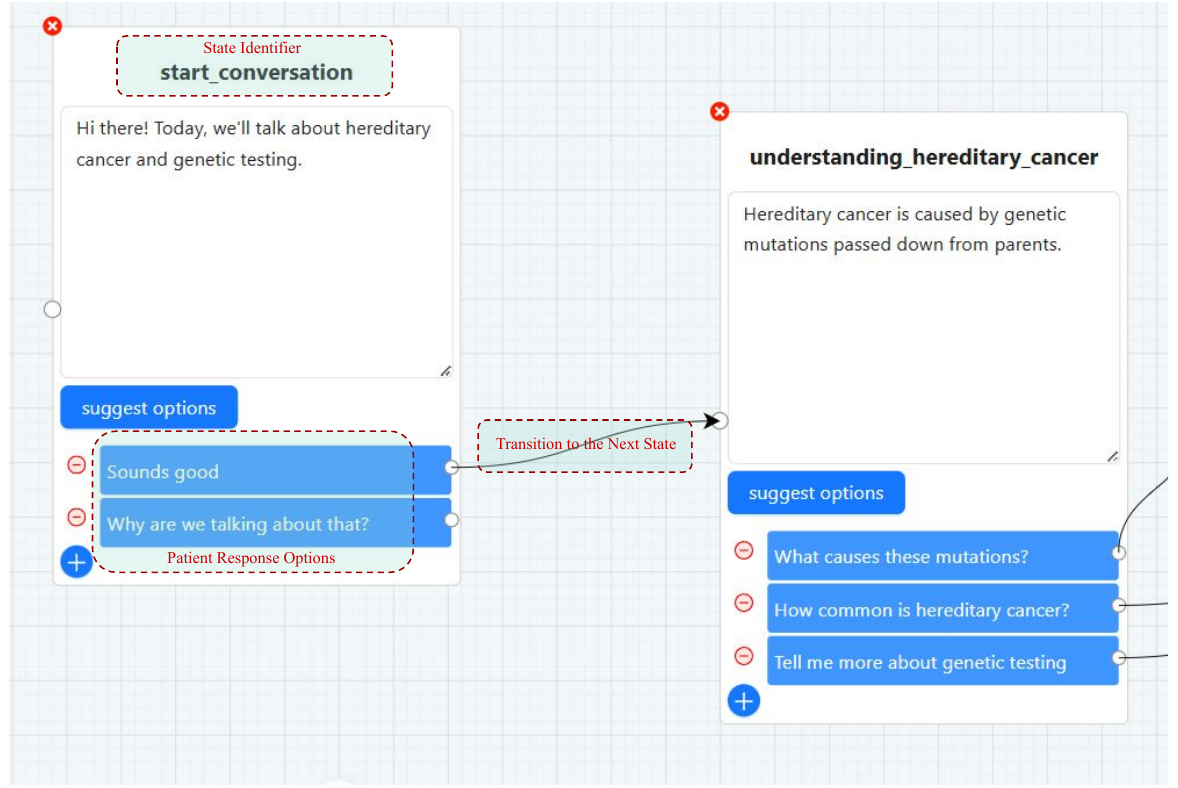}
  \caption{The visual representation of a dialogue state, as part of the dialogue finite state machine in HealthDial}
  \Description{A visual representation of a dialogue state, as part of the dialouge finite state machine}
  \label{fig:fsm}
\end{figure}

HealthDial consists of two main components: 1) a session planning phase that takes the educational material as text, and creates a nested list of session topics and in-session subtopics, and 2) a conversation flow generation phase that takes the educational material, and the session-based plan from the previous phase, and creates a dialogue for each session using a simple Behavior Markup Language (BML) that only includes textual utterances. When creating the dialogue, HealthDial assumes a structured, agent-initiated dialog, in which each discourse segment \cite{grosz1986attention} is comprised of an agent’s utterance to the patient, and the patient’s response to the utterance. Dialogues are modeled using finite state machines, where a state refers to a pair consisting of a virtual agent utterance, uttered in the virtual agent interface through voice or text, and its corresponding patient response options, represented in the interface as buttons that a patient can click on or tap. During the interaction, when a patient chooses a response option, they are directed to a unique next state, determined by a state transition, or connection. The visual representation of these concepts within HealthDial, as seen by the dialogue author, can be viewed in \autoref{fig:fsm}.

\subsection{Implementation Details}
\begin{figure*}
  \centering
  \includegraphics[width=\linewidth]{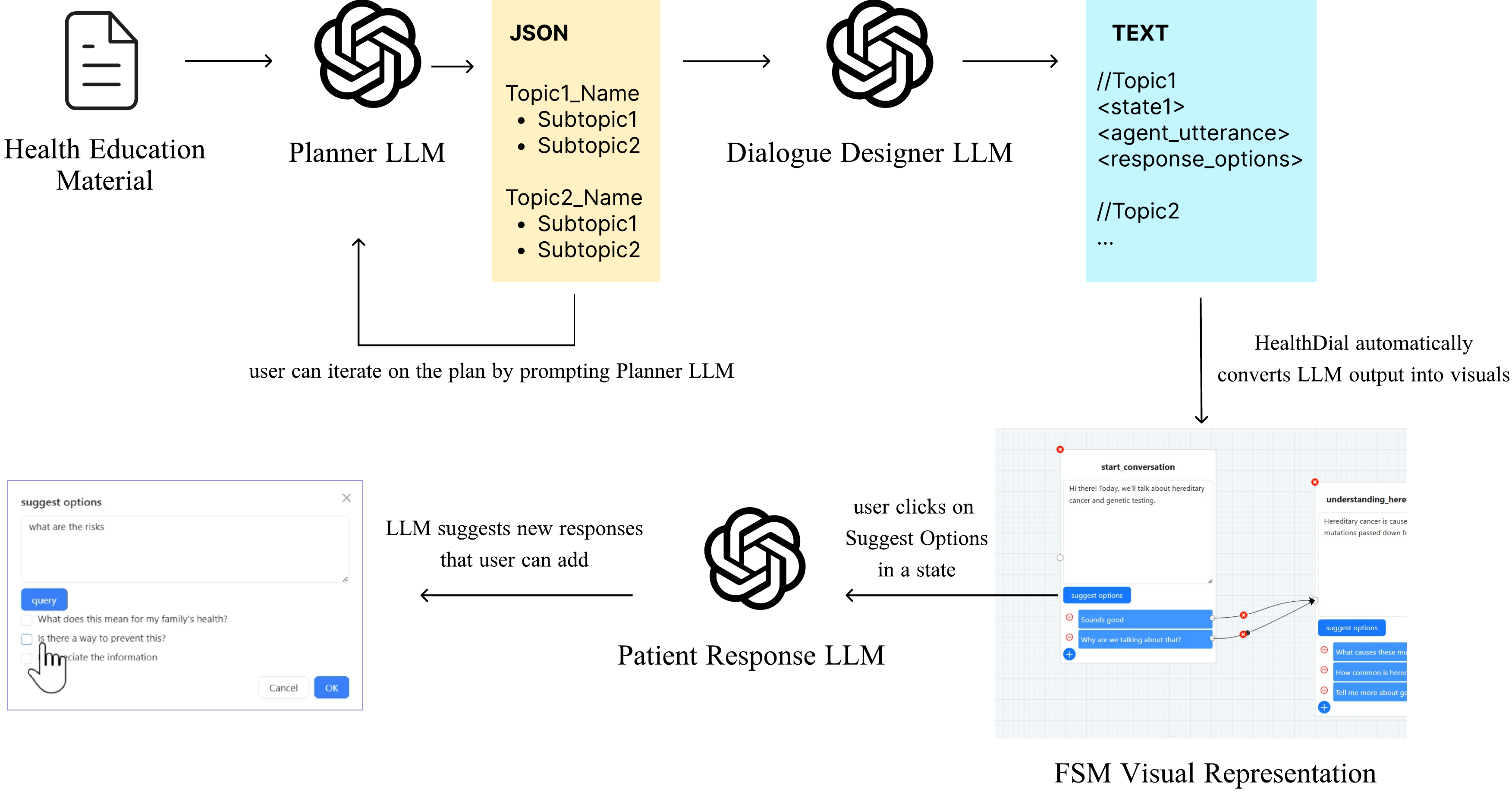}
  \caption{HealthDial System Architecture and LLM orchestration}
  \Description{Figure 2. HealthDial System Architecture and LLM Orchestration}
  \label{fig:system}
\end{figure*}

HealthDial was developed as a responsive and interactive authoring environment\footnote{Software developed using the Vue.js framework} for healthcare professionals and educators with varying levels of technical expertise. The system was developed as a locally-stored application and is structured into multiple modular components that guide users through the process of transforming health education materials into validated, session-based finite state machine (FSM) dialogue flows suitable for delivery via intelligent virtual agents. \autoref{fig:teaser} shows the user journey when using HealthDial to create conversations for healthcare virtual agents, and \autoref{fig:system} provides a glance into the backend working machanisms, including the LLM orchestration\footnote{All LLMs used in our system are instances of OpenAI GPT-4o}. The prompts used to direct the three LLMs in the system can be found in our anonymous repository \footnote{https://anonymous.4open.science/r/HealthDial-09BC/}.

\paragraph{Input and Material Preprocessing}
Upon accessing HealthDial, users are presented with an initial landing page where they can supply health education content either by directly pasting text or importing text files. This flexible intake ensures compatibility with a range of established health education resources and documentation workflows.

\paragraph{Topic Extraction and Session Planning}
Once the input material is loaded, users initiate the topic derivation process by activating the "Convert to Topics" function. This triggers a call to a backend service invoking an API to the Planner LLM. The model is prompted with the uploaded educational material and, where applicable, user directions for adjusting coverage or emphasis. The LLM returns a structured JSON plan, decomposing the content into an ordered list of sessions, each with a unique topic and an associated set of key educational points. This planning step is interactive; users may iteratively refine session topics and key points by providing revision cues to the LLM, ensuring alignment with instructional intent and clinical guidelines.

\paragraph{Conversational Flow Generation}
After approval of the session-based plan, users proceed to the conversation authoring phase. By selecting "Convert Topic to Conversations," an invocation to the Dialogue Designer LLM processes the original material alongside the approved session plan, generating an initial draft FSM for each session. The FSMs are output using a hierarchical, human-readable BML modeled after hierarchical transition networks previously established for virtual agent dialogue specification \cite{bickmore2022intelligent}. Each FSM encodes agent utterances, possible patient response options, and deterministic transitions between dialogue states.

Within the HealthDial interface, these FSMs are automatically rendered into dynamic visual diagrams. Dialogue states are represented as labeled nodes containing editable agent utterance fields; patient-selectable patient options are displayed as subordinate text rectangles. State transitions are visualized as directed arrows. This design provides full transparency and granular editability, reducing the risk of LLM hallucinations influencing patient-facing agent behavior. While the current implementation does not focus on nonverbal agent behaviors (e.g., gestures, facial expressions), our format is extensible. Nonverbal cues can be incorporated by extending the markup language and prompt templates, leveraging LLMs for gesture suggestion or enabling manual annotation of behavior tags.

\paragraph{Interactive Editing and Versioning}
Users may double-click to directly edit any agent or patient utterance, add or delete states and transitions, or reorder conversational paths. A dedicated “Suggest Options” button at each state enables users to invoke a third API, Patient Response LLM. This model, conditioned on the session context and current state, offers contextually appropriate additional patient response options. This supports rapid expansion or refinement of dialogue interactivity in response to author feedback.

Sessions—represented as discrete topics—can be globally managed from a navigation pane. Users can add, remove, or reorder topics, with deletions cascading to their associated dialogue FSMs.

\paragraph{Export and Integration}
Once dialogue flows across all sessions are finalized, authors can export their work using the “Export” function. This compiles all FSMs—reflecting all manual and LLM-assisted revisions—into a machine-readable text file using the established FSM markup format. This output is intended for direct deployment within virtual agent delivery platforms, ensuring that authored content is transparent, auditable, and shielded from unsafe spontaneous LLM output at runtime.

In summary, HealthDial’s system design tightly integrates LLM-powered content generation and structured human-in-the-loop editing, producing virtual agent dialogue that is both scalable and robust. The use of FSMs and a visual, no-code editing interface provides authors with comprehensive control and flexibility, while minimizing potential for unsafe or inaccurate agent communication.

\section{Feasibility Study}
To test the usability and feasibility of the proposed system, we conducted a mixed-method study in which individuals with health counseling knowledge used HealthDial for an example task of hereditary cancer genetics education. We chose this task because of its clear objectives and educational nature, which makes it suitable for a conversational interaction where a virtual agent provides knowledge, and patient response options include particular questions or clarifications that a patient may need. Our study was approved by the IRB of our institution. Here, the term "AI" refers to the orchestrated group of LLMs within HealthDial, which gave participants suggestions for a session-based plan, the dialogues for all sessions, and the patient response options.

\subsection{Recruitment}
To get insights into how acceptable this approach of creating virtual agents can be, we recruited individuals with health counseling experience through social media at our institution. During screening, we ensured that participants had at least a few previous counseling sessions or health conversations, to gain relevant feedback when they used HealthDial.

\subsection{Task Description} \label{sec:task}
The task assigned to participants was presented as follows: "\textit{Imagine you are a genetic counselor and are hoping to teach your patients about hereditary cancer screening. You are told that you can create an automated virtual counselor that delivers that conversation, and think of possible responses or questions that the patient may have. You are given this tool to help you create that conversation. Once you are done, we will embed your conversation in a virtual counselor so you can test what you made}". For the knowledge participants were tasked to disseminate, we used an educational pamphlet on hereditary cancer screening, developed by the Wilmot cancer institute at University of Rochester \cite{wilmot}. The pamphlet (which we are unable to present due to licensing) includes a description of cancer, its genetic and environmental causes, and highlights the importance of screening for preventing the disease. The material also touches upon the emotional and legal aspects of screening test results. Participants were allowed to copy and use any part of the pamphlet within the tool, while also typing their own preferred text.

\subsection{Study Procedure}
The study took a total of 75 minutes. After consent, the baseline questionnaire was administered. Participants were then given an overview of the task (Section \ref{sec:task}), and were introduced to the HealthDial application. The researchers then provided a short tutorial of how the system works, and answered participant questions about the format and meaning of various elements (e.g., states, transitions, patient response options, and agent utterances). Then, participants were given 5 minutes to review the pamphlet, and were asked to begin performing the task of creating the conversation, with a maximum of 30 minutes to spend.

Once a participant felt their task was completed or the 30-minute duration was over, a research assistant exported their designed conversation into a format usable by a virtual agent system. The new file format included the content designed by the participant, as well as nonverbal expressions and gestures automatically generated using BEAT Animation Toolkit \cite{cassell2001beat}. Participants spent another 5 minutes testing their own conversations, this time acting as a patient receiving information via the virtual agent system.

Post-interaction questionnaire was then administered, followed by the final semi-structured interview to learn about the participant's experience. Participants were then thanked and compensated for their time.

\subsection{Measures}
We describe the measures used in our study at various stages. A breakdown of the outcomes measures is also shown in \autoref{tab:measures}, and measure titles will be used for brevity in reporting outcomes.
\paragraph{Baseline Survey}
Participants were first asked to fill out a baseline survey, where they responded about their socio-demographic background, including their occupation and prior counseling experience. The depth of their counseling expertise, and their attitudes toward use of computers and AI for both every-day and health-related activities were measured using 7-point Likert questions. Additionally, participants responded to similar Likert scales to indicate their perceived need for a tool that they can use to plan their health conversations. 

\paragraph{Post-Interaction Survey}
During participants' interaction with the HealthDial authoring tool, we recorded 1) the time they took to finish the task of finalizing their dialogue state machine, and 2) the number of revisions they made within the application, to edit the LLM suggestions (including topic, dialogue content, and response option changes). We did not count participants' manipulation of transition connections, or any reverted changes, as these are often essential steps that follow other content changes. After finishing the task followed by interacting with the virtual agent, the post interaction survey was administered. Participants first responded to the System Usability Scale (SUS) \cite{brooke1996sus}, a standard scale widely used to measure usability of software systems. Self-reported task difficulty, perceived time taken to finish the task, satisfaction with HealthDial, the virtual agent interaction, and the resulting health conversation were then measured using single-item 7-point Likert scales. With similar scales, participants were also asked about their perception of HealthDial's helpfulness in planning, their intent in using the tool for both virtual agent dialogue design and for planning their own conversations, and how much they would rely on HealthDial for genetic counseling and other topics. Questions specific to the LLM suggestions were then asked, using measuring the AI's helpfulness in the authoring task, their trust in the AI suggestions, and how much they felt the suggestions were aligned with their own counseling practice. Lastly, participants responded to the PEMAT scale \cite{shoemaker2014development}, a standard assessment tool for patient health education materials. We used a short version of this questionnaire, which contains 13 items with binary \textit{Agree}/\textit{Disagree} scores (1/0), convertible into two percentage scores of \textit{Understandability} and \textit{Actionability} to describe health media.

\paragraph{Semi-Structured Interviews}
Participants were also interviewed by the research assistant with more in-depth questions about their experience with the HealthDial authoring tool, the virtual agent interaction, AI suggestions within HealthDial, and their intent to use the tool if it were available for health education. A more detailed list of interview questions can be found in \autoref{sec:interview}  

\section{Results}
\subsection{Study Sample}
We recruited 9 students and employees of our institution (6F, 2M, 1 undisclosed; 2 Caucasian, 7 Asian). Participant backgrounds ranged from medicine, pharmacy and dentistry, to education and health technology. When asked for their expertise or depth of previous exposure to counseling at baseline (7-point Likert scale with anchors [Complete Novice - Complete Expert], all participants rated themselves higher than 3, with an average rating of 6.1 across all participants. 

All participants identified as regular users of computers, with 2 rating themselves as experts in this regard. When asked whether they often use computers to plan health conversations, the average results were slightly more frequent than \textit{Sometimes} on a 7-point Likert scale with anchors [Not at All - Very Often] (Med=4, IQR=3). A trending Wilcoxon Signed Rank test revealed that participants indicated a frequent need for computer tools that could help them with planning conversations (Z=1.8, p=0.06; Med=5, IQR=2). 

When asked about how often they used AI in work activities, a median rating of 6 (IQR=3) was found. Additionally, answers to the question regarding their attitudes toward AI had a median rating of 8 (IQR=3), with both scores showing that participants were generally favoring and accustomed to AI applications for everyday work tasks. That said, median of ratings for the question of how often participants used AI to plan their health conversations was aligned with the neutral point \textit{Sometimes} on a 7-point Likert with anchors [Not at All - Very Often] (Med=4, IQR=2) possibly showing a lower trust in AI decision making for health applications.

\subsection{Quantitative Outcomes}
We conducted statistical quantitative analyses on both standardized multi-item and single-item measures across usability, subjective experience, and attitudes toward AI suggestions. This section reports the quantitative findings from our feasibility and usability study with counselors and students.

We began our analyses with Shapiro-Wilk tests of normality to enable choice of descriptives and statistical tests for ratio and composite measures of SUS and PEMAT. For single-item measures, we use Median/IQR for descriptives and performed non-parametric Wilcoxon Signed Rank tests to compare 1) ratings from the neutral midpoint of each item, and 2) pre- vs. post interaction measures. 


\paragraph{Usability and Perceived Experience}
The mean SUS score for HealthDial was 79 (SD = 11.1, Min = 62.5, Max = 97.5), indicating good usability according to benchmarks established by the interaction design community \cite{interactiondesignorg}. For general task experience, we found that \textbf{Tool Satisfaction} (Z=2.3, p<0.05), \textbf{Tool Help Plan} (2.7, p<0.01), \textbf{Intent to Use Tool} (Z=2.3, p<0.05), \textbf{Need A Tool} (Z=2.7, p<0.01), \textbf{Need This Tool} (Z=2.2, p<0.05) were all rated significantly higher than the neutral of 4 on their respective Likert scale. Ratings for \textbf{Intent to Develop Agent} using the authoring tool were trending higher than the neutral midpoint (Z=1.8, p=0.06). 
These results indicate high perceived value and user satisfaction for HealthDial as a conversation planning and authoring platform. However, the same test for \textbf{Agent Satisfaction} was non-significant, with a Med(IQR) of 5(3). Notably, the \textbf{Intent to Develop Agent} using HealthDial trended higher than neutral (Z = 1.8, p = 0.06), signaling potential uptake for broader virtual agent authoring beyond the study context.

\paragraph{Attitudes Toward LLM Suggestions}
Participants consistently rated the AI-generated scaffolding and suggestions highly. Specifically, on 7-point Likert items, \textbf{AI Helpfulness} (Z=2.5, p<0.05), \textbf{AI Trust} (Z=1.9, p<0.05), \textbf{AI Alignment} (Z=2.0, p<0.05) were all significantly higher than the neutral of 4, providing evidence that the orchestrated LLMs within HealthDial were seen as trustworthy sources of brainstorming and content generation, and aligned with participants' professional expectations. That said, participants' confident \textbf{Reliance} on HealthDial for this and other topics, had scores non-signifcantly higher than neutral, with a Med(IQR) of 5(3) and 5(2), respectively. Additionally, the objective measure of number of revisions in LLM suggestions was low across participants (Med=4, IQR=4), showing a general acceptance of AI suggestions.

\paragraph{Task Challenge}
For \textbf{Task Difficulty} and the \textbf{Perceived Time} required to complete the authoring task, Wilcoxon Signed Rank tests showed no significant deviation from the neutral scale point, suggesting that participants experienced the task as neither unduly challenging nor time-consuming. This reinforces the SUS findings and supports HealthDial’s usability for non-expert users. Lastly, the objective measure of the time participants took had a mean of 24.6 and SD of 5.2 seconds.

\paragraph{Shifts in Perceived Need for Authoring Tools}
Pre- and post interaction comparisons reflected an increase in participants’ perceived need for structured digital tools to plan health education conversations. Specifically, a related-samples Wilcoxon Signed Rank test comparing baseline \textbf{Need Computer – Health} ratings with post-interaction ratings of both \textbf{Need A Tool} and \textbf{Need This Tool} revealed a significant increase following HealthDial use for Need A Tool (Z = 2.3, p < 0.05),
and a trending increase for \textbf{Need This Tool} (Z = 1.8, p = 0.06). These shifts suggest that hands-on experience with HealthDial made participants more aware of the usefulness of dedicated authoring tools for health-related conversations.

\paragraph{Health Communication Quality}
One Sample Wilcoxon Signed Rank tests on single-item measure of \textbf{Conversation Quality} (Med=5, IQR=2) had non-significant results when comparing to the neutral midpoint. To further assess the quality of the generated dialogues, we analyzed participants' judgments of the final conversation based on \textbf{PEMAT} assessment \cite{shoemaker2014development}. The dialogues were rated highly for both understandability and actionability (\textbf{Understandability} M = 96.3, SD = 11.1; \textbf{Actionability} M = 92.6, SD = 14.7). These findings support the tool’s general effectiveness in facilitating the creation of patient-facing materials that are clear and actionable, meeting guidelines for health communication.

\subsection{Qualitative Outcomes}
We conducted thematic analysis of participant interviews following their use of HealthDial to author and deploy multi-session health education virtual agent dialogues. Below, we present major themes and their relationship with the design of our system.

\paragraph{Virtual Agent Delivery of Health Education}
Most participants considered the use of virtual agents for patient education promising, emphasizing their potential in delivering accessible, consistent health information outside clinical encounters. As P4 noted, "\textit{for patients to have some sort of knowledge without stepping into a doctor office...I felt was nice}". Multiple participants highlighted that a virtual agent could reduce the cognitive and emotional load on patients, particularly before, between, or following clinical visits. The approach was seen as particularly valuable in populations with limited health literacy or access, as P8 described: "\textit{A lot of times people have a hard time explaining what is happening to them...this tool would really give edge to that because you can just go and tell this virtual counselor what is happening and get routed to the right care}". A caregiver at nursing homes, P8 also mentioned: "\textit{This is really good for older people because they don't know...you have to give them the information over multiple times}", highlighting the agent’s accessibility benefits for older populations. The agent's voice-based delivery of health education for those who may be visually impaired, and its potential for integration in waiting rooms or home use were commonly endorsed ways to increase reach of care through the use of virtual agents.


\paragraph{LLM Suggestions: Trust, Helpfulness and Time Savings}
Overall, participants expressed high levels of trust in the LLM-generated suggestions as scaffolding their authoring process—particularly when it came to coverage and breadth. "\textit{It was like four times in a row that AI would just smash it out of the park...kept being exactly what I had in mind. And that was exciting}" [P3]. Many described the initial AI draft as "\textit{a path, I would say. I know what way to go...It gave you some direction}" [P5].
Nevertheless, human intervention remained essential—not only for verification but for tailoring content to specific patient groups, cultural considerations, and local context. "\textit{I had the opportunity to make changes...according to what I wanted}" [P4]. The value was articulated especially for users with less expertise in counseling: "\textit{If you don't have the knowledge, you’d have a much harder time with a blank slate}" [P5].

\paragraph{Avoiding Jargon and Supporting Personalization}
Participants repeatedly emphasized the importance of lay-friendly language, highlighting HealthDial’s utility in reducing medical jargon from source materials: “\textit{The pamphlet had some medical terms but this made it more straightforward…like everyday language}" [P8]. P3, who volunteers for a local addiction support center, reported having fine control for linguistic register and sensitive tailoring: "\textit{I wanted to use colloquial urban language throughout but avoid using culturally sensitive languages, and the AI did that really well}". That said, P6 had ideas for further tailoring of language to various groups, through defining various personas of patients within HealthDial: "\textit{maybe there could be a better way where you ask AI to generate possible questions that a child might ask, a senior might ask, or someone well versed in medicine might ask. And then...if you have a child who's using the agent, I would be able to select something so that it's interacting with them while knowing that it should be presenting the child's anticipated questions, and present their answers}".

\paragraph{Constrained Input vs. Openness: Tradeoffs in Dialogue Authoring}
The FSM approach to dialogue modeling, with its emphasis on predefined patient options, drew mixed reactions. Some participants appreciated the safety and predictability: "\textit{It gives the person designing this more control...they know whatever possible answers there are that will be given}" [P6], while others wanted a more open-ended experience: "\textit{[The patient would] run out of options when the agent is saying something and [they] have something else in mind, but just have to choose between [the options]}" [P2]. Voice or text-based free input was frequently suggested as a way to not only increase flexibility, but also use textual patient inputs as new ideas for patient response options later on---albeit with concern for safety and validation in health contexts. 


\paragraph{Anticipation and Iteration: Covering Patient Questions} Many participants highlighted the challenge of anticipating a broad range of patient questions, indicating both the strengths and limitations of AI support. While HealthDial's automatic suggestion of patient response options often sparked new ideas ("\textit{[The response options] were unique. I didn’t have it in my mind, but then when I saw it, it made sense.}" [P2]), some participants described difficulty in achieving coverage of possible patient responses due to the (modality and attention) differences between the experience as an author and the conversation actually playing out in the virtual agent system: "\textit{I was planning my questions by myself...but when I sat on [the] patient side...I realized something is missing}" [P1]. As a solution, a more iterative authoring paradigm, i.e., previewing agent behavior, then refining, was suggested to bridge initial gaps: "\textit{If you have any options in the authoring [tool] that you can just play it out to listen to what things are there}" [P1]. The value of iteration was brought up multiple times, so that the dialogue author can feel confident in their own design. 


\paragraph{Supporting Novices and Experienced Counselors through Empowerment and Practice}
Participants from both medical and non-medical backgrounds reported the tool fostered empowerment, particularly for those less confident in health communication: "\textit{For people who are probably nervous, this tool might be a good thing because you can practice your conversations}" [P4]. Reflecting on their practice, some participants highlighted a lack of proper training for novice clinicians: "\textit{As beginner dentists we often had a more senior doctor that we shadowed and tried to learn how to talk to patients from them...If we had this tool it would be much easier}" [P2]. More experienced providers sometimes saw less need for preparation, but appreciated the ability to "\textit{practice play}", prepare for difficult conversations and "pre-concoct rebuttals" [P3], and regarded it as a significant advance over being thrown into the deep end of conversations without preparation. HealthDial was also recognized as a generalizable authoring approach, with participants readily imagining use beyond cancer genetics. P6, a former school counselor, highlighted: "\textit{I could see using it to anticipate what questions children or parents might have and how I might respond to each}". These findings suggest that even though HealthDial was primarily created to support virtual agent development, it may also be effective at encouraging healthcare providers to practice and prepare for their own conversations.

In summary, our findings show that HealthDial was deemed a promising, flexible dialogue authoring system. Participants acknowledged that the system achieves a balance between automatic scaffolding and human refinement, supports clarity and topic coverage, and empowers creators from diverse backgrounds—albeit with identified needs for more flexible conversational structures.

\section{Discussion}
Our study demonstrates that HealthDial successfully lowers the barriers for healthcare professionals—regardless of their technical background—to author understandable and effective dialogues for virtual health agents. By bringing a no-code, LLM-assisted interface into dialogue authoring, participants were able to move from static health materials to interactive, FSM-based conversational flows with surprising ease and high levels of trust in the AI-generated suggestions. Importantly, users reported that HealthDial helped ensure comprehensive topic coverage while preserving their agency over clinical nuance and individual tailoring. This reflects a broader trend in virtual agent technology to empower domain experts by translating technical complexity into intuitive, editable workflows \cite{jordan2001tools}.

A central finding is that HealthDial enables a paradigm shift from the historically high-dependency collaboration between health experts and technical designers toward a more democratized, independent authoring model for health virtual agents. In our study, participants valued HealthDial’s deterministic FSM output since every conversational branch could be reviewed, revised, and validated, thus safeguarding patient safety and shielding agents from LLM hallucinations and unsafe advice \cite{bickmore2021mitigating, beinema2022wool}. The blend of LLM-driven creativity and strict author control responds to longstanding safety and coverage concerns with conversational agent design in clinical contexts \cite{laranjo2018conversational, denecke2023framework}.

Our work also highlights the role of AI scaffolding in enhancing the efficiency and the quality of the authoring process. Participants described the AI-drafted conversational outlines as a "path" that facilitated brainstorming and minimized blank-slate anxiety. The system’s suggestions for patient responses not only sparked new ideas but also promoted a more thorough anticipation of patient queries—an area traditionally cited as a shortcoming even in role-play-based authoring \cite{pilnick2018using, rossen2012crowdsourcing}. Furthermore, the iterative, visual nature of HealthDial supports a "preview-then-refine" model, empowering creators to incrementally improve dialogue clarity, coverage, and personalization, and echoing constructionist educational theories which argue that teaching material (here, by constructing agent dialogue) enhances the author’s own understanding \cite{halan2014virtual}.

An emergent opportunity identified in our results is the dual role of HealthDial as both an agent creation tool and a training/practice platform for clinicians. Participants found value in using HealthDial to anticipate "difficult conversations," practice responses, and organize complex health content, essentially using the design process as a rehearsal for real interactions. This form of simulated practice could support novice providers in developing communication confidence and skills, as has been validated in recent patient simulation, peer-tutoring, and virtual human research \cite{halan2014virtual, chu2024synthetic, mukherjee2024polaris, albright2018using, steenstra2025scaffolding}.

Finally, HealthDial demonstrates a high degree of generalizability. The combination of LLMs for initial scaffolding, deterministic FSMs for safe deployment, and human-in-the-loop editing provides a robust model for authoring across a wide range of health education topics and for diverse patient populations \cite{beinema2022wool}. Moreover, the system’s architecture is compatible with integration in both clinical practice and broader health literacy efforts, such as pre-visit education and remote counseling scenarios \cite{bickmore2013randomized}.

\section{Limitations \& Future Work}
While the results from our pilot study are encouraging, several limitations should be noted. Our participant pool was relatively small, and disproportionately composed of young, AI-favoring users; future studies must involve a broader demographic, especially experienced genetic counselors and those with more varied attitudes toward AI to ensure ecological validity.
The goal of our study was to evaluate the usability and general acceptability of the tool, motivating the study of a single group of participants. That said, future work can involve an experiment to understand the role of various system features, such as by adding a control group that does not include any AI assistance within the tool. In addition, objective quality metrics such as real-world patient comprehension or behavioral outcomes could provide a more rigorous evaluation beyond self-report scales and usability indices. 
Finally, while HealthDial achieved strong subjective scores for usability and content quality, further work is warranted to optimize onboarding, support open-ended interactions, and facilitate iterative refinement and validation of patient response options similar to prior work \cite{choi2020leveraging, rossen2012crowdsourcing}. The current implementation was locally deployed; a web-based version could better respond to real-world conditions and accessibility needs. Additionally, we are currently woring towards an end-to-end approach that allows for independent dialogue authoring and agent deployment, eliminating the need to convert HealthDial output into machine-readable formats as a separate step. These improvements will be crucial for realizing HealthDial’s full potential in diverse health education and clinical care settings.

\section{Conclusion}
HealthDial represents a promising advance in the authoring of patient-facing health dialogues by uniting LLM-enabled drafting with safety-oriented, no-code editing interfaces. Our feasibility study indicates that HealthDial is perceived as usable, trustworthy, and beneficial both for virtual agent development and as a tool for preparing clinicians for complex health conversations. Participants successfully translated static health education content into interactive agent dialogues, benefitted from LLM-suggested scaffolding, and felt empowered to customize and validate every aspect of the conversational flow. Given these findings, HealthDial lays important groundwork for democratizing the creation of healthcare virtual agents and scaling patient education efforts, while maintaining crucial expert oversight.


\bibliographystyle{ACM-Reference-Format}
\bibliography{references}

\appendix

\section{Measure Descriptions}
\begin{table*}
  \caption{Outcome Measures used in our feasibility study, along with the stage at which they were administered. The abbreviation "pt" refers to the number of points in a Likert scale.}
  \label{tab:measures}
  \begin{tabular}{cccl}
    \toprule
    Title & Items(type) &Description& Anchors\\
    \midrule
    SUS (Usability)  & 10(5-pt)& Brooke et al. \cite{brooke1996sus}& Disagree-Agree\\
    Task Difficulty  & 1(7-pt)& Difficulty of creating this conversation&Too Easy-Too Hard\\
    Perceived Time  & 1(7-pt)& Time taken for creating this conversation&Too Short-Too Long\\
    Tool Satisfaction  & 1(7-pt)& Satisfaction with authoring tool & Not at All-Very Much\\
    Tool Help Plan  & 1(7-pt)& Authoring tool would help plan my conversations& Not at All-Definitely\\
    Intent to Use Tool  & 1(7-pt)& Would use the tool for my conversation planning&Strongly Disagree-Strongly Agree\\
    Conversation Quality  & 1(7-pt)& Quality of the final dialogue met expectations& Not at All-Very Much\\
    Agent Satisfaction  & 1(7-pt)& Satisfaction with agent's delivery of the dialogue& Not at All-Very Much\\
    Intent to Develop Agent & 1(7-pt)& Would use authoring tool to develop agents& Strongly Disagree-Strongly Agree\\
    Need A Tool  & 1(7-pt)& Need for health-planning computer tool& Strongly Disagree-Strongly Agree\\
    Need This Tool  & 1(7-pt)& Need for health-planning computer tool like this& Strongly Disagree-Strongly Agree\\
    Reliance-This Topic  & 1(7-pt)& Would rely on authoring tool for this topic& Would Never-Always\\
    Reliance-Other Topics  & 1(7-pt)& Would rely on authoring tool for other topics& Would Never-Always\\
    AI Helpfulness  & 1(7-pt)& AI suggestions helpful in authoring& Not at All-A Lot\\
    AI Trust  & 1(7-pt)& AI suggestions trusted in authoring& Not at All-Very Much\\
    AI Alignment  & 1(7-pt)& AI suggestions aligned with practice in authoring& Not at All-Very Much\\
    PEMAT Understandability  & 10(2-pt)& Shoemaker et al. \cite{shoemaker2014development}& Agree/Disagree\\
    PEMAT Actionability  & 3(2-pt)& Shoemaker et al. \cite{shoemaker2014development}& Agree/Disagree\\
    \bottomrule
  \end{tabular}
\end{table*}

\section{Interview Guide} \label{sec:interview}
\begin{itemize}
    \item Main impressions of the tool? What did you like or dislike about the tool?
    \item Was there any point you felt stuck
    \item How could the tool have been more helpful to you? Did you wish anything was different / added / removed?
    \item What were your impressions of the virtual agent delivering this material?
    \item As a counselor, do you imagine this authoring tool could be useful in your health education conversations? How? 
    \item Do you feel this virtual agent delivery method may be helpful to you? In what circumstances?
    \item How clear / understandable / actionable did you feel the information ended up to be?
    \item Would you feel comfortable relying on a similar system for other health education topics?
    \item Were the smart suggestions impacting you in any way? How? What did you like/dislike about the AI suggestions?
    \item Did you trust the information provided by the AI?
    \item How easy was it to use the AI in the tool? 
    \item Did you feel the AI was hurting the process in any way?
    \item Did you have a particular group of patients in mind when designing the conversation? How relevant was the final conversation to that group, in your opinion?

\end{itemize}

\end{document}